\documentclass[10pt,letterpaper]{article}
\usepackage{opex3}
\usepackage{graphicx}  

\begin{document}

\title{11\,W narrow linewidth laser source at 780nm for laser cooling and manipulation of Rubidium}
\author{S.~S. San\'e, S. Bennetts, J.~E. Debs, C.~C.~N. Kuhn, G.~D. McDonald,\\P.~A. Altin, J.~D. Close and N.~P. Robins}
\address{Quantum Sensors Lab, Department of Quantum Science,\\Australian National University, Canberra, 0200}
\email{nick.robins@anu.edu.au} 
\homepage{http://atomlaser.anu.edu.au} 

\begin{abstract}
We present a narrow linewidth continuous laser source with over 11\,W output power at 780\,nm, based on single-pass frequency doubling of an amplified 1560\,nm fibre laser with 36\% efficiency. This source offers a combination of high power, simplicity, mode quality and stability. Without any active stabilization, the linewidth is measured to be below 10\,kHz. The fibre seed is tunable over 60\,GHz, which allows access to the D$_2$ transitions in $^{87}$Rb and $^{85}$Rb, providing a viable high-power source for laser cooling as well as for large-momentum-transfer beamsplitters in atom interferometry. Sources of this type will pave the way for a new generation of high flux, high duty-cycle degenerate quantum gas experiments.
\end{abstract}


\section{Introduction}

The rapid progress of atomic physics over the past few decades has largely hinged upon the development of high power, narrow linewidth laser sources for manipulating and probing atoms. Increased laser power allows for higher flux and collection efficiency in magneto-optical traps \cite{klempt}, as well as further improvements in lattice-based cooling techniques \cite{weiss}. In atom interferometry, narrow linewidth, high power near-resonant lasers are a prerequisite to achieving higher sensitivities with large-momentum-transfer beamsplitting \cite{holger24, markgrav, stuart}.

The most common atomic species used for making alkali gas Bose-Einstein condensates (BECs) is $^{87}$Rb. This is due in part to its excellent scattering properties which allow efficient evaporative cooling, but also because of the availability of inexpensive 780nm CD burner laser diode sources which can be stabilized using an external cavity grating \cite{wieman,hansch}. Today, for a modest cost, a grating stabilised diode master laser/tapered amplifier configuration will produce a cw source with a 10's of \,kHz linewidth and up to 2\,W of output power. However, the spatial mode quality is typically poor, leading to usable powers of $<500$\,mW after spatial filtering.   These power levels and linewidths are not adequate for next generation high flux Rb BEC and interferometers and thus an alternate approach is required.

In recent years a great deal of effort has focused on the development of high power narrow linewidth CW sources at 589 nm primarily for sodium guide star applications, but also for laser cooling.  Currently doubling raman fibre lasers in an external resonant cavity have demonstrated powers of 50W \cite{calia1} and doubling efficiencies of 86\% at 25W \cite{calia2}. Another approach by Chiow et al. used a modified Coherent 899 Ti:sapphire laser to achieve 6W of light at 852 nm by injection locking. Frequency stabilization to a high-finesse optical cavity resulted in a linewidth of $<$ 1 kHz \cite{holgertisaph}. This technology allowed for efficient high-order Bragg diffraction in the largest area atom interferometer produced to date \cite{big}. Diode pumped alkali vapour lasers have now demonstrated 48W in Cs \cite{knize}.

At the 780nm Rb wavelength, tunable Ti:sapphire and Alexandrite lasers have demonstrated up to 6W, offering the option of injection locking \cite{morris}. Zweiback {\em et. al} have demonstrated 28W from a diode pumped alkali vapour laser \cite{krupke}. By cascading two PPLN crystals, Thompson et al. were able to obtain 20\% SHG efficiency and generate 900mW of light at 780 nm \cite{thompson}.  Recently the atomic physics community has also begun to investigate the possibility of using compact cw frequency-doubled sources for portable atom interferometry based sensors \cite{lienhart,bouyer}. 

In this paper we present a compelling argument to move to a doubled laser system in atomic physics labs working with Rubidium. We have produced a 11.4\,W cw laser at 780.24\,nm with a 6\,kHz linewidth by single-pass frequency doubling in a single PPLN cystal. The doubling efficiency is 36\%. The setup is simple and robust, relying on a narrow-linewidth fiber laser \cite{therock} to provide a highly stable seed, and a low noise 30\,W fibre amplifier \cite{IPG} to generate the high powers required for efficient doubling.

\section{Apparatus}
The experimental setup is shown in Figure 1. The source is an amplifed NP Photonics Rock fibre laser with a centre wavelength of 1560.48\,nm and a tuning range of $\pm30\,$GHz. Only $\pm150$\,MHz tuning is readily available via piezo control, otherwise the laser must be tuned with temperature. This laser has a specified linewidth of $<5$\,kHz integrated over 100\,ms. The output frequency appears nearly insensitive to acoustic noise, particularly compared to an external cavity diode laser. We do observe a slow thermal drift of the seed laser, which is easily counteracted by a low bandwidth servo loop with an error signal provided by saturation spectroscopy of Rubidium with the doubled light. The 1560.48\,nm seed is amplified by an IPG Photonics fiber amplifier with a maximum output power of 30\,W. The amplified beam has a $1/e^2$ diameter of 1.1\,mm and is linearly polarized.

\begin{figure}
\centering\includegraphics[width=\textwidth]{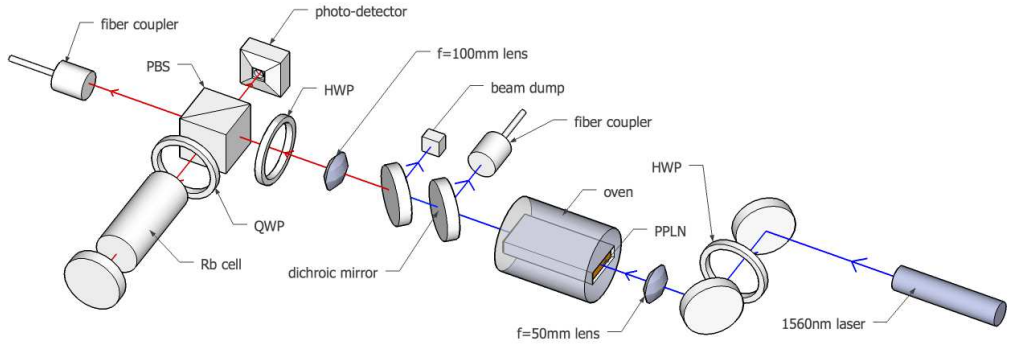}
\label{fig:apparatus}
\caption{Schematic of the experimental setup. The seed and fiber amplifier are not shown in the diagram. PPLN - periodically-poled lithium niobate crystal, HWP - half-wave plate, QWP - quarter-wave plate, PBS - polarizing beamsplitter.  After the oven, the 780nm and 1560nm light are separated by the dichoric mirrors. }
\end{figure}

Using a single plano-convex lens with a 50\,mm focal length, this beam is focused into the centre of a 40\,mm long periodically-poled lithium niobate (PPLN) cystal. The crystal is 1\,mm thick and has five 1\,mm wide gratings, each with a different domain period \cite{crystal}. In this work we have used a grating with a 19.5\,$\mu$m domain period. The crystal is housed inside a temperature-stabilized oven on a three-axis translation stage, and held at 81.60\,$^\circ$C for optimal phase matching. The polarisation of the light incident on the crystal is controlled using a $\lambda/2$ waveplate, which is optimized to achieve maximum doubling efficiency.

The linearly polarised 780\,nm light exiting the crystal is filtered by two dichroic mirrors and collimated with a 100\,mm lens.  This light is analyzed via saturated absorption spectroscopy using rubidium vapor, which also provides the locking signal for the seed laser (a fibre modulator is used to generate the necessary frequency sidebands for locking). The remaining 1560\,nm light reflected by the dichroic mirrors can be used for dipole trapping. The optical setup is robust and compact, and does not require any active control of the optical components, with the exception of temperature-stabilizing the PPLN crystal.

\section{Results}
The power in the second harmonic is plotted as a function of input power in Figure 2. A maximum efficiency of 36\% is achieved, giving 11.4\,W of output power at 780\,nm. These data were taken without any adjustment of the crystal temperature or alignment. This efficiency compares very favourably with more complex and lower power cavity-enhanced doubling systems at these wavelengths \cite{doub}. The inset in Figure 2 shows the spatial mode of the 780\,nm light, after collimation from the doulber.  By tuning the seed laser temperature and piezo, we can scan through all of the rubidium D$_2$ transitions without a noticeable change in power.  

\begin{figure}[htbp]
\centering\includegraphics[width=10cm]{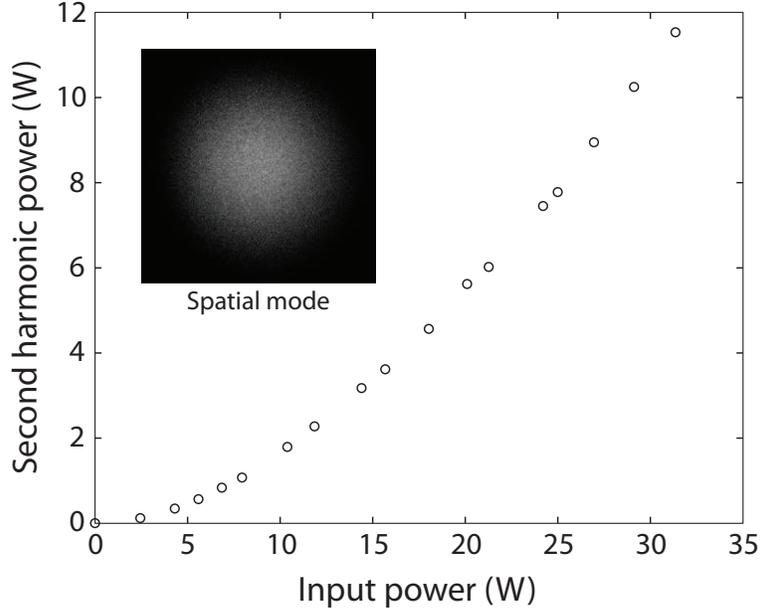}
\label{fig:efficiency}
\caption{Measured second harmonic power as a function of input power from a single 40\,mm PPLN crystal. The maximum output power is 11.4\,Watts at 780\,nm. The inset shows the spatial mode of the output.}
\end{figure}

To measure the linewidth of the 780\,nm beam, a small portion of output power is directed through a fiber-coupled electro-optic phase modulator (modulation frequency 50\,MHz) and then into an acoustically isolated unequal path length Mach-Zehnder inteferometer. One arm of the interferometer is passed through a 10\,m single-mode optical fibre, giving a fringe spacing of 21\,MHz. The output of the interferometer is monitored on a fast photodetector, and demodulation and low-pass filtering is used to generate an error signal which can be straightforwardly calibrated to the fringe spacing. The frequency noise spectrum obtained at the zero-crossing of the error signal is given in Figure 3. Above 10\,kHz, the measurement is limited by detector and electronic noise, aside from a peak at 500\,kHz which corresponds to a noise feature in the seed laser. By integrating the noise spectrum over the range plotted in Figure 3, we determine the linewidth of the frequency-doubled light to be 6\,kHz over 100\,ms, which is comparable to that of the NPP seed laser specified as $<5$\,kHz.

\begin{figure}[htbp]
\centering\includegraphics[width=10cm]{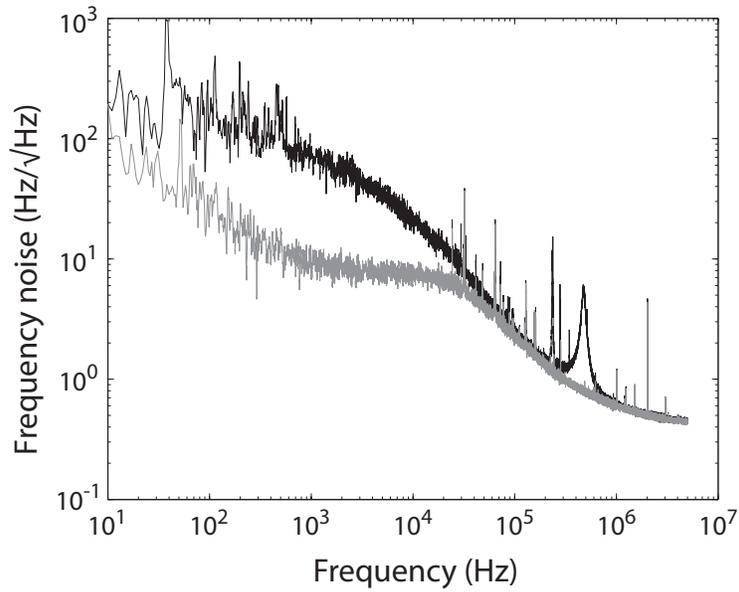}
\label{fig:spectrum}
\caption{Frequency noise spectrum measured using an unequal path length Mach-Zehnder interferometer as described in the text. The gray curve shows the detector noise. Integrating from 10\,Hz to 5\,MHz gives a linewidth of 6\,kHz.}
\end{figure}

\section{Conclusion}
We have presented a frequency-doubled laser source at 780\,nm, which provides over 11\,W of continuous power in a high quality Gaussian mode with a linewidth of 6\,kHz integrated over 100ms. We believe this to be the highest reported power at this wavelength.  The system does not require locking to a high finesse cavity or using multiple crystals as in previous efficient frequency doubling experiments, which makes the apparatus robust and remarkably simple to set up. This system has applications in a variety of atomic physics experiments, including for high-order Bragg diffraction in precision atom interferometry.

\section{Acknowledgements}

The authors would like to thank Tim Lam for his assistance with the linewidth measurement and Nikita Simakov for useful discussions.  NPR thanks Nina and Alexander Robins for experimental support work.  Specific product citations are for the purpose of clarification only and are not an endorsement by the authors or the ANU.  This work was supported in part by the Australian Research Council Discovery program.


\begin{thebibliography}{99}

\bibitem{klempt} S. Jollenbeck, J. Mahnke, R. Randoll, W. Ertmer, J. Arlt, and C. Klempt. ``Hexapole-compensated magneto-optical trap on a mesoscopic atom chip", Phys. Rev. A 83, 043406 (2011).
\bibitem{weiss} M. Olshanii and D. Weiss, ``Producing Bose-Einstein condensates using optical lattices", Phys. Rev. Lett. 89, 090404 (2002).
\bibitem{holger24} H. Muller, S. Chiow, Q. Long, S. Herrmann, and S. Chu. ``Atom interferometry with up to 24-photon-momentum-transfer beam splitters", Phys. Rev. Lett. 100, 180405 (2008). 
\bibitem{markgrav} S. Dimopoulos, P. W. Graham, J. M. Hogan, M. A. Kasevich, and S. Rajendran. ``Atomic gravitational wave interferometric sensor", Phys. Rev. D 78, 122002 (2008).
\bibitem{stuart} S. S. Szigeti, J. E. Debs, J. J. Hope, N. P. Robins, and J. D. Close. ``Why momentum width matters for atom interferometry with Bragg pulses", New J. Phys. 14, 023009 (2012).
\bibitem{wieman}K. B. MacAdam, A. Steinbach, and C. Wieman.  ``A narrow-band tunable diode laser system with grating feedback, and a saturated absorption spectrometer for Cs and Rb", Am. J. Phys. 60, 1098 (1992).
\bibitem{hansch}L. Ricci, M. Weidem\"uller, T. Esslinger, A. Hemmerich, C. Zimmermann, V. Vuletic, W. K\"onig and T.W. H\"ansch. ``A compact grating-stabilized diode laser system for atomic physics", Opt. Comm. 117, 541 (1995).
\bibitem{calia1}L. R. Taylor, Y. Feng, and D. B. Calia, ``50W CW visible laser source at 589nm obtained via frequency doubling of three coherently combined narrow-band Raman fibre amplifiers", Optics Express, Vol. 18, 8540-8555 (2010).
\bibitem{calia2} Y. Feng, L. R. Taylor, and D. B. Calia, ``25 W Raman-fiber-amplifier-based 589 nm laser for laser guide star", Optics Express, Vol. 17, 19021-19026 (2009).
\bibitem{holgertisaph} S. Chiow, S. Herrmann, H. Muller, S. Chu. ``6 W, 1 kHz linewidth, tunable continuous-wave near-infrared laser", Opt. Express 17, 5246 (2009).
\bibitem{big} S.-Y. Lan, P.-C. Kuan, B. Estey, P. Haslinger, and H. M\"uller. ``Influence of the Coriolis force in atom interferometry", accepted Phys. Rev. Lett. (2012).
\bibitem{knize} B.V. Zhdanov, J. Sell, R.J.Knize, ``Multiple laser diode array pumped Cs laser with 48W output power", Electronics Letters, 44, 582 - 583 (2008).
\bibitem{morris} J. Walling, O. Peterson, R.  Morris, ``Tunable CW alexandrite laser",
IEEE Journal of Quantum Electronics, 16, 120-121 (1980).
\bibitem{krupke} J. Zweiback and W. F. Krupke
``28W average power hydrocarbon-free rubidium diode pumped alkali laser"
Optics Express, 18, 1444-1449 (2010).
\bibitem{thompson}R. J. Thompson, M. Tu, D. C. Aveline, N. Lundblad, L. Maleki., ``High power single frequency 780nm laser source generated from frequency doubling of a seeded fiber amplifier in a cascade of PPLN crystals", Opt. Express 11, 1709 (2003).
\bibitem{lienhart}F. Lienhart, S. Boussen, O. Carat, N. Zahzam, Y. Bidel, and A. Bresson. ``Compact and robust laser system for rubidium laser cooling based on the frequency doubling of a fiber bench at 1560 nm", Appl. Phys. B 89, 177 (2007).
\bibitem{bouyer}V. M\'enoret, R. Geiger, G. Stern, N. Zahzam, B. Battelier, A. Bresson, A. Landragin, and P. Bouyer. ``Dual-wavelength laser source for onboard atom interferometry", Opt. Lett. 36, 4128 (2011).
\bibitem{therock}  NP Photonics, The Rock.  
\bibitem{IPG} IPG photonics. 
\bibitem{crystal} PPLN crystal supplied by Covesion Ltd.
\bibitem{doub} J. Feng, Y. Li, X.Tian, J. Liu, and K. Zhang. ``Noise suppression, linewidth narrowing of a master oscillator power amplifier at 1.56nm and the second harmonic generation output at 780nm", Opt. Express 16, 11871 (2008).

\end{thebibliography}
\end{document}